\documentclass{jetpl}
\usepackage{graphicx}

\twocolumn \lat
\begin{document}

\title{Pinning of spiral fluxons by giant screw dislocations in
$YBa_2Cu_3O_{7-\delta}$ single crystals: Josephson analog of the
fishtail effect} \rtitle{} \sodtitle{Josephson analog of the
fishtail effect}

\author{S. Sergeenkov$^{a}$, L.Jr. Cichetto$^{a}$, V.A.G. Rivera$^{a}$,  C. Stari$^{a,b}$,
E. Marega$^{c}$, C.A. Cardoso$^{a}$, F.M.
Araujo-Moreira$^{a}$\/\thanks{Research Leader, e-mail:
faraujo@df.ufscar.br}}

\address{$^{a}$Departamento de F\'{i}sica e Engenharia
F\'{i}sica, Grupo de Materiais e Dispositivos, Centro
Multidisciplinar para o Desenvolvimento de Materiais Cer\^amicos,
Universidade Federal de
S\~ao Carlos, S\~ao Carlos, SP, 13565-905 Brazil\\
$^{b}$Instituto de F\'{i}sica, Facultad de Ingenieria, Julio
Herrera y Reissig 565, C.C. 30, 11000, Montevideo, Uruguay\\
$^{c}$ Instituto de F\'{i}sica, USP,  S\~ao Carlos, SP, 13560-970
Brazil}
\date{\today}

\abstract{ By using a highly sensitive homemade AC magnetic
susceptibility technique, the magnetic flux penetration has been
measured in $YBa_2Cu_3O_{7-\delta}$ single crystals with giant
screw dislocations (having the structure of the Archimedean
spirals) exhibiting $a=3$ spiral turnings, the pitch $b=18.7\mu m$
and the step height $c=1.2nm$ (the last parameter is responsible
for creation of extended weak-link structure around the giant
defects). The magnetic field applied parallel to the surface
enters winding around the weak-link regions of the screw in the
form of the so-called spiral Josephson fluxons characterized by
the temperature dependent pitch $b_f(T)$. For a given temperature,
a stabilization of the fluxon structure occurs when $b_f(T)$
matches $b$ (meaning an optimal pinning by the screw dislocations)
and manifests itself as a pronounced low-field peak in the
dependence of the susceptibility on magnetic field (applied
normally to the surface) in the form resembling the high-field
(Abrikosov) fishtail effect. }


\maketitle

{\bf 1. Introduction.} Probably one of the most intriguing
phenomena in the vast area of superconductivity is the so-called
fishtail effect (FE) which manifests itself as a maximum in the
field dependence of the magnetization. There are many different
explanations regarding the existence and evolution of FE (see,
e.g.,~\cite{1,1a,2,3,4,4a,5,6,7,7a,8,9} and further references
therein). However, the most plausible scenario so far is based on
the oxygen deficiency mediated pinning of Abrikosov vortices in
non-stoichiometric materials with the highest possible $T_C$ due
to a perfect match between a vortex core and a defect size. It has
been unambiguously demonstrated that by diminishing the deficiency
of oxygen in undoped $YBa_2Cu_3O_{7-\delta}$, the FE decreases and
tends to disappear completely. A characteristic field at which
this phenomenon usually occurs is of the order of a few Tesla.

In this Letter we report on observation of a low-field (Josephson)
analog of the high-field (Abrikosov) FE in $YBa_2Cu_3O_{7-\delta}$
single crystals with giant screw dislocations (having the
structure of Archimedean spirals), which we attributed to the
perfectly matched pinning of the spiral $2D$ Josephson fluxons
with the temperature dependent spiral pitch $b_f(T)$ (predicted by
Aranson, Gitterman, and Shapiro~\cite{9a}) by the Archimedean
spirals with the pitch $b=18.7\mu m$.
\begin{figure}
\centerline{\includegraphics[width=8cm,angle=0]{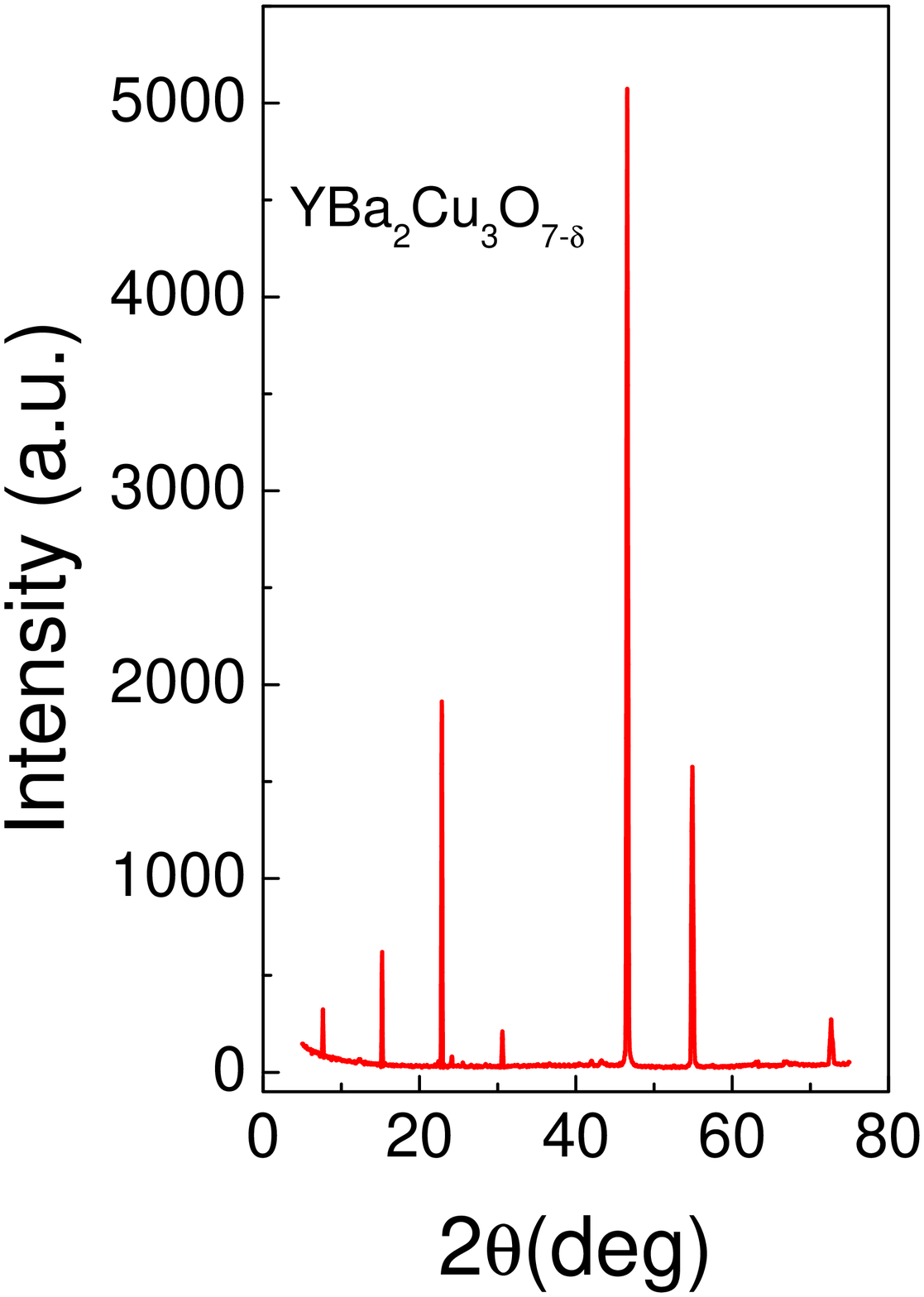}}
\vspace{0.5cm} \caption{Fig.\ref{fig:fig1}. XRD pattern of
$YBa_2Cu_3O_{7-\delta}$ single crystals. } \label{fig:fig1}
\end{figure}
\begin{figure}
\centerline{\includegraphics[width=8cm,angle=0]{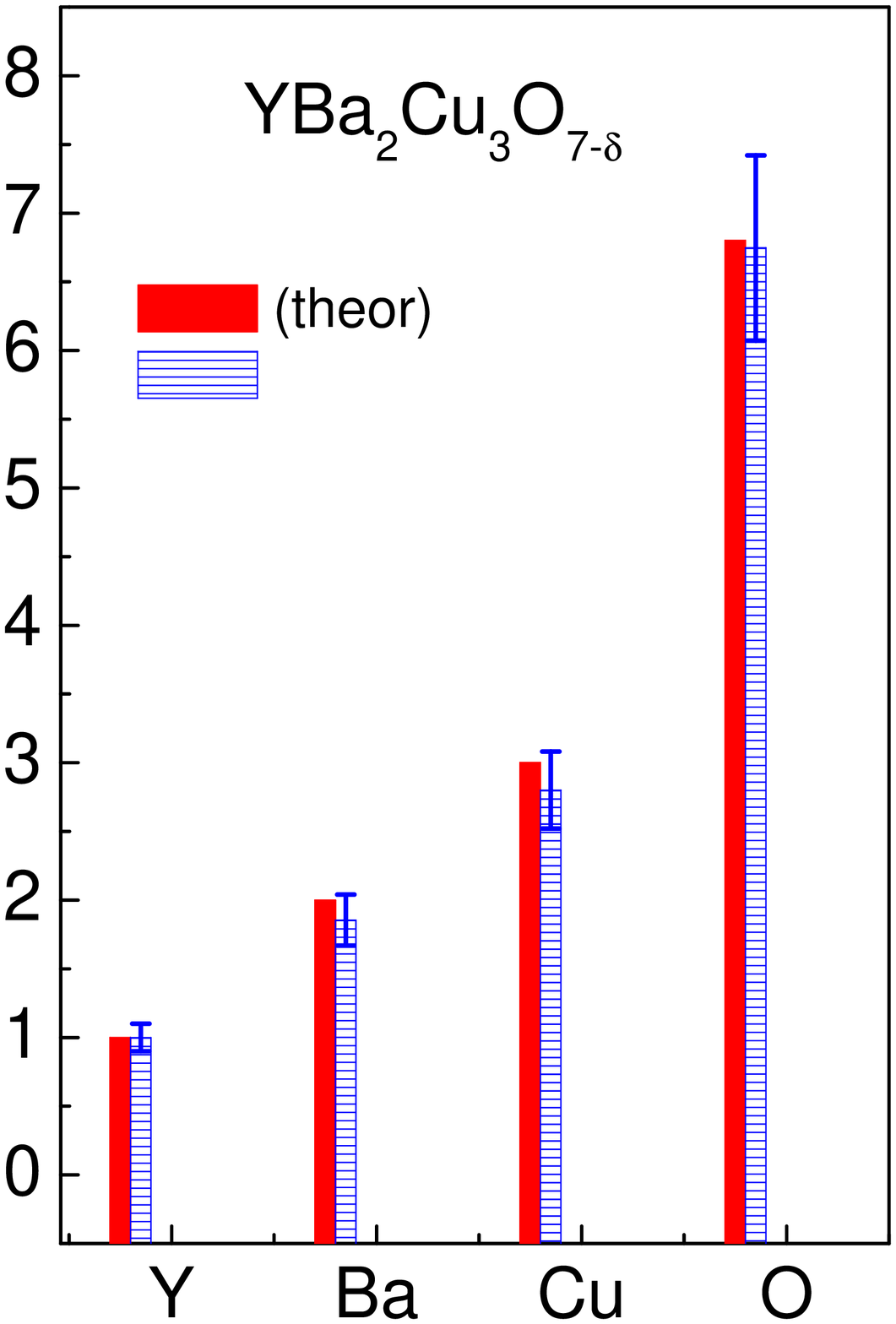}}
\vspace{0.5cm} \caption{Fig.\ref{fig:fig2}. EDX spectrum of
$YBa_2Cu_3O_{7-\delta}$ single crystals prepared under argon
atmosphere along with the theoretical value for each component
(error bars are taken with the accuracy of $\pm 10 \%$). }
\label{fig:fig2}
\end{figure}

\begin{figure}
\centerline{\includegraphics[width=8.0cm,angle=0]{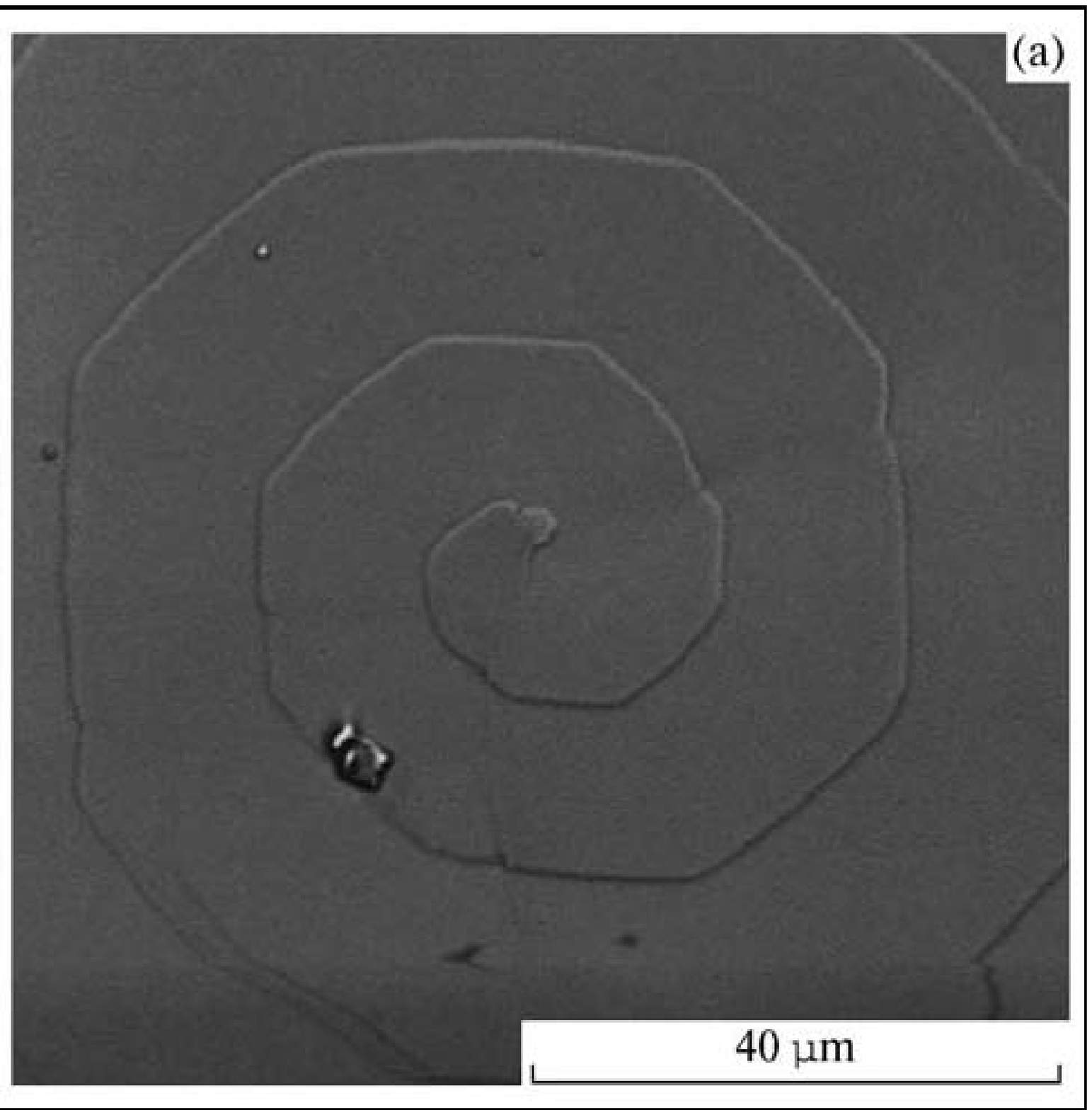}}\vspace{0.0cm}
\centerline{\includegraphics[width=8.0cm,angle=0]{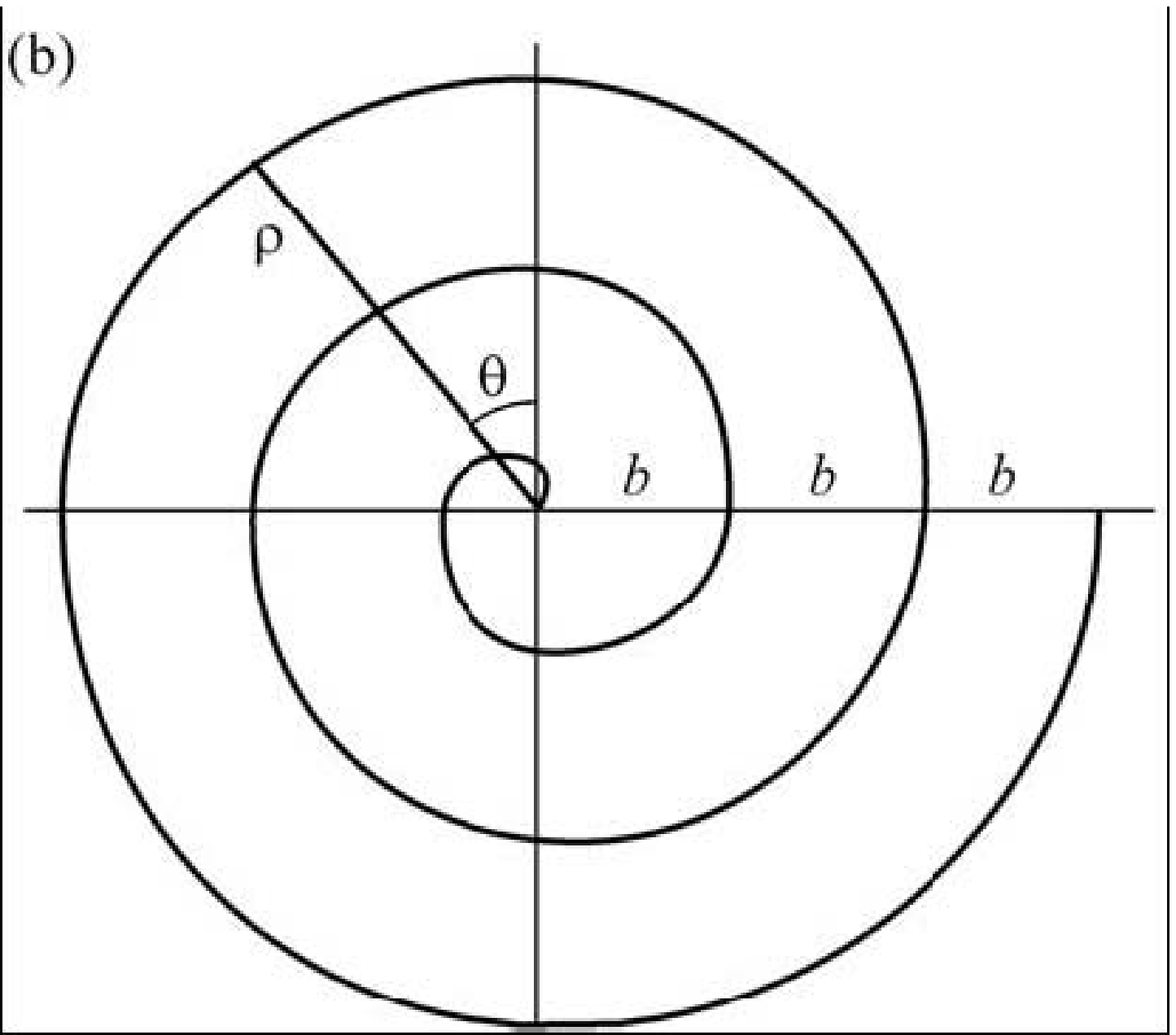}}
\vspace{0.5cm} \caption{Fig.\ref{fig:fig3}. (top) SEM photography
of $YBa_2Cu_3O_{7-\delta}$ single crystals showing a giant screw
dislocation (having the structure of the Archimedean spiral) with
$a=3$ spiral turnings and the pitch $b=18.7\mu m$ (magnification
$500$ times). (bottom) Sketch of the one-armed Archimedean spiral
with three $360^{\circ}$ turnings (counterclockwise) in polar
coordinates $(\rho, \theta)$ and a pitch (distance between
successive turnings) $b$.} \label{fig:fig3}
\end{figure}

{\bf 2. Sample Characterization.} One of the most reliable growing
techniques used to obtain high-quality crystalline superconducting
samples is related to the so-called chemical route based on the
self-flux method~\cite{10}. It allows to obtain samples with
almost no secondary phases (as compared to the traditional method
of solid-state reaction~\cite{11}). Our samples ($2.8\times 2.1
\times 0.9mm^3$) were prepared following the self-flux method with
different routes using both the heat treatment and environmental
conditions (oxygen and argon atmospheres)~\cite{12}. The phase
purity and the structural characteristics of our samples were
confirmed by both scanning electron microscopy (SEM) with Energy
Dispersive  X-ray Analysis (EDX) accessory and X-ray diffraction
(XRD), including the standard Rietveld analysis. Crystalline
phases were confirmed by XRD  using  a Rigaku instrument with a
$Si$ $111$ monochrometer. XRD pattern (shown in
Fig.\ref{fig:fig1}) was taken using $CuK\alpha$ radiation
($\lambda = 1.5418$\AA). Quantitative analysis of elements present
in both crucibles and crystals was evaluated with accuracy of up
to $0.01 \%$.  The resulting diagram is shown in
Fig.\ref{fig:fig2} which depicts the relative value of the
components present in single crystal (normalized with respect to
$Y$) along with the theoretically predicted estimates for $\delta
=0.2$. Fig.\ref{fig:fig3} shows  SEM surface scan topography of
$YBa_2Cu_3O_{7-\delta}$ single crystals with a giant screw
dislocation having the structure of the Archimedean spiral on
$100\mu m$ scale (with $a=3$ spiral turnings and the pitch
$b=18.7\mu m$) along with a schematic structure of the one-armed
Archimedean spiral (with three $360^{\circ}$ turnings
counterclockwise) which in polar coordinates is defined by the
equation $\rho=b\theta$ where $b$ is a pitch (distance between
successive turnings).

\begin{figure}
\centerline{\includegraphics[width=8cm]{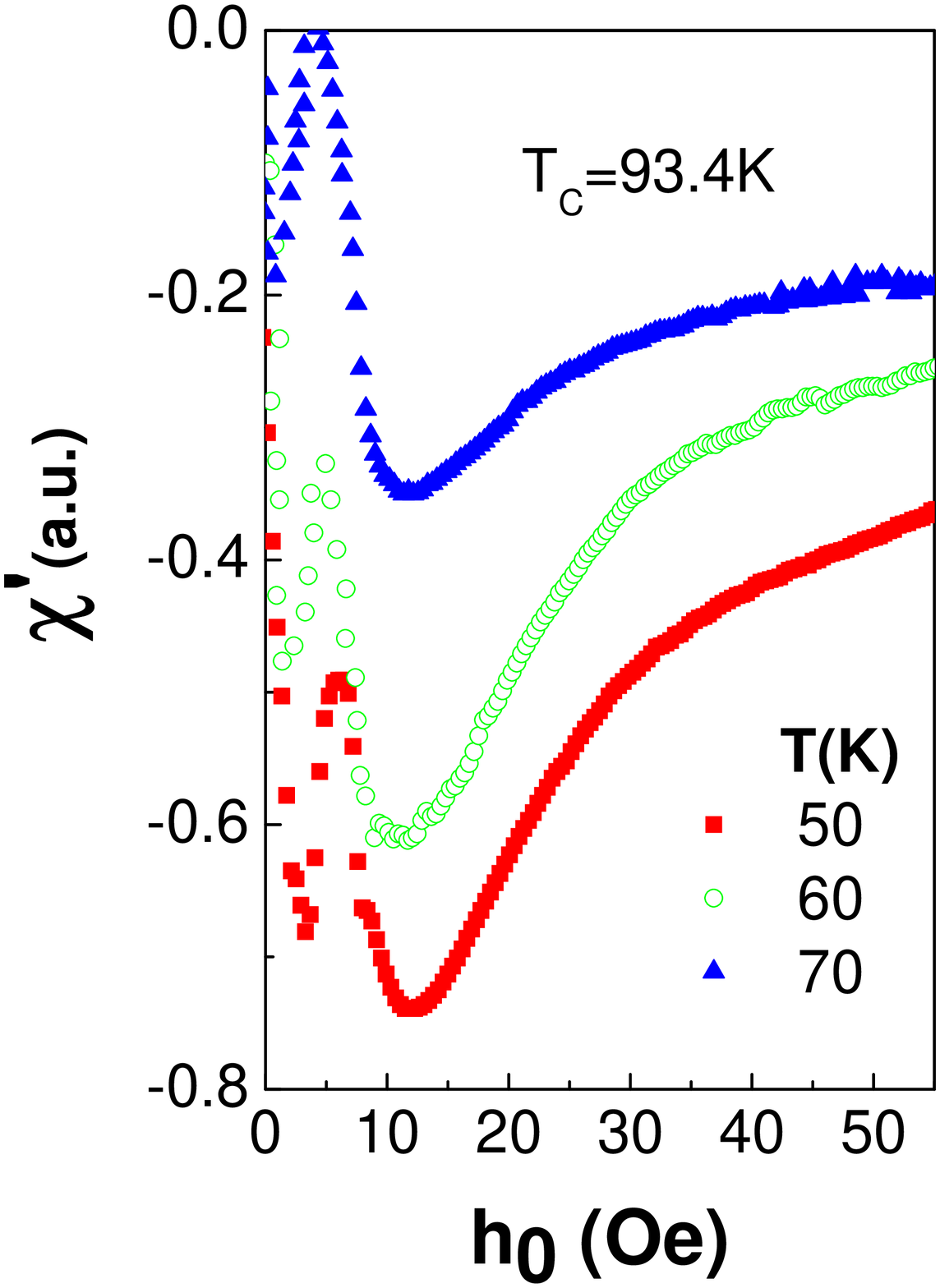}}\vspace{0.5cm}
\caption{Fig.\ref{fig:fig4}. A fishtail-like structure of magnetic
field dependence of the AC susceptibility $\chi^{'}(T,h_0)$ in
$YBa_2Cu_3O_{7-\delta}$ single crystals for various temperatures.}
\label{fig:fig4}
\end{figure}

{\bf 3. Results and Discussion.} AC susceptibility technique has
proved very useful and versatile tool for probing low-field
magnetic flux penetration in superconductors on macroscopic level
(see, e.g.,~\cite{13,13a,13b,13c,14} and further references
therein). In this paper, AC measurements were made by using a
high-sensitivity homemade susceptometer based on the screening
method and operating in the reflection configuration~\cite{13}. We
have measured the complex AC susceptibility $\chi_{ac}=\chi^{'}+i
\chi^{''}$ as a function of the applied AC field $h_{ac}(t)=h_0
\cos(\omega t)$ (with the amplitude $0.01Oe\le h_0\le 50Oe$ and
frequency $1kHz \le \omega \le 30kHz$) taken at fixed temperature.
Initially, to ensure the formation of the so-called one-armed
spiral fluxons~\cite{9a} winding around the weak-link regions of
the Archimedean spirals, $h_{ac}$ was applied parallel to the
sample's surface. To measure the response from the pinned
(stabilized) fluxons, the magnetic field was applied normally to
the surface. Fig.\ref{fig:fig4} shows a typical dependence of the
AC susceptibility $\chi^{'}(T,h_0)$ on AC field amplitude $h_0$
for different temperatures along with the corresponding critical
temperature, $T_C$. It is important to notice that a rather high
value of $T_C$ in our single crystals is achieved by preparing
samples in argon atmosphere. Notice an unusual reentrant-type
behavior of $\chi^{'}(T,h_0)$ along with a clear peak-like
structure (around $h_0=5Oe$) shifting to lower fields with
temperature (which indeed strongly resembles the high-field
Abrikosov FE and by analogy will be called a low-field Josephson
FE).

Turning to the discussion of the obtained results, let us show how
the formation and pinning (by the Archimedean spirals) of spiral
Josephson vortices (fluxons) can manifest itself as a pronounced
low-field peak in the measured magnetic response. According to
this scenario, the peak field $h_p(T)$ (where susceptibility
exhibits a fishtail-like low-field maximum) can be related to the
fluxon critical field $h_f(T)=\Phi_0/db_f$. The latter defines the
onset for fluxon motion. More precisely, for $h_0>h_f(T)$ spiral
fluxons start to rotate along the weak-link regions of the
Archimedean spirals. Here, $d=2\lambda_L+c \simeq 2\lambda_L$ is
the width of the contact with $\lambda_L$ being the London
penetration depth of the superconducting region and $c$ the
thickness of non-superconducting region in the vicinity of the
screw defined by the vertical step height (distance between
terraces of the screw dislocation along the $c$-axis of $YBCO$
crystal). Typically~\cite{14a}, $c$ is of the order of a unit cell
(that is $c\simeq 1.2nm$). $b_f(T)$ is the temperature dependent
pitch of the spiral Josephson vortex. According to the theoretical
predictions~\cite{9a}, $b_f(T)=5\lambda_J(T)/\gamma(T)$, where
$\lambda_J=\sqrt{\Phi_0/ \mu_0J_Cd}$ is the Josephson penetration
depth (which also defines a size of the conventional $1D$
Josephson vortex and the lower Josephson field
$h_J=\Phi_0/\lambda_Jd$) with $J_C(T)$ being the critical current
density; the temperature dependent factor $\gamma(T)$ accounts for
the onset border for spiral fluxon motion~\cite{9a}. In our case,
$\gamma(T)=h_0/h_f(T)>1$. Recall~\cite{9a} that the dynamics of
the spiral fluxons is parametrically defined in polar coordinates
as follows, $x_f(t)=\rho \cos[n\theta(\rho)-\Omega t]$ and
$y_f(t)=\rho \sin[n\theta(\rho)-\Omega t]$ where $n$ is the number
of arms and $\Omega$ is the angular velocity of fluxon rotation.
In particular, for $\theta(\rho)=\rho /b_f$, the above relations
describe the Archimedean spiral with pitch $b_f$.
\begin{figure}
\centerline{\includegraphics[width=7.2cm,angle=0]{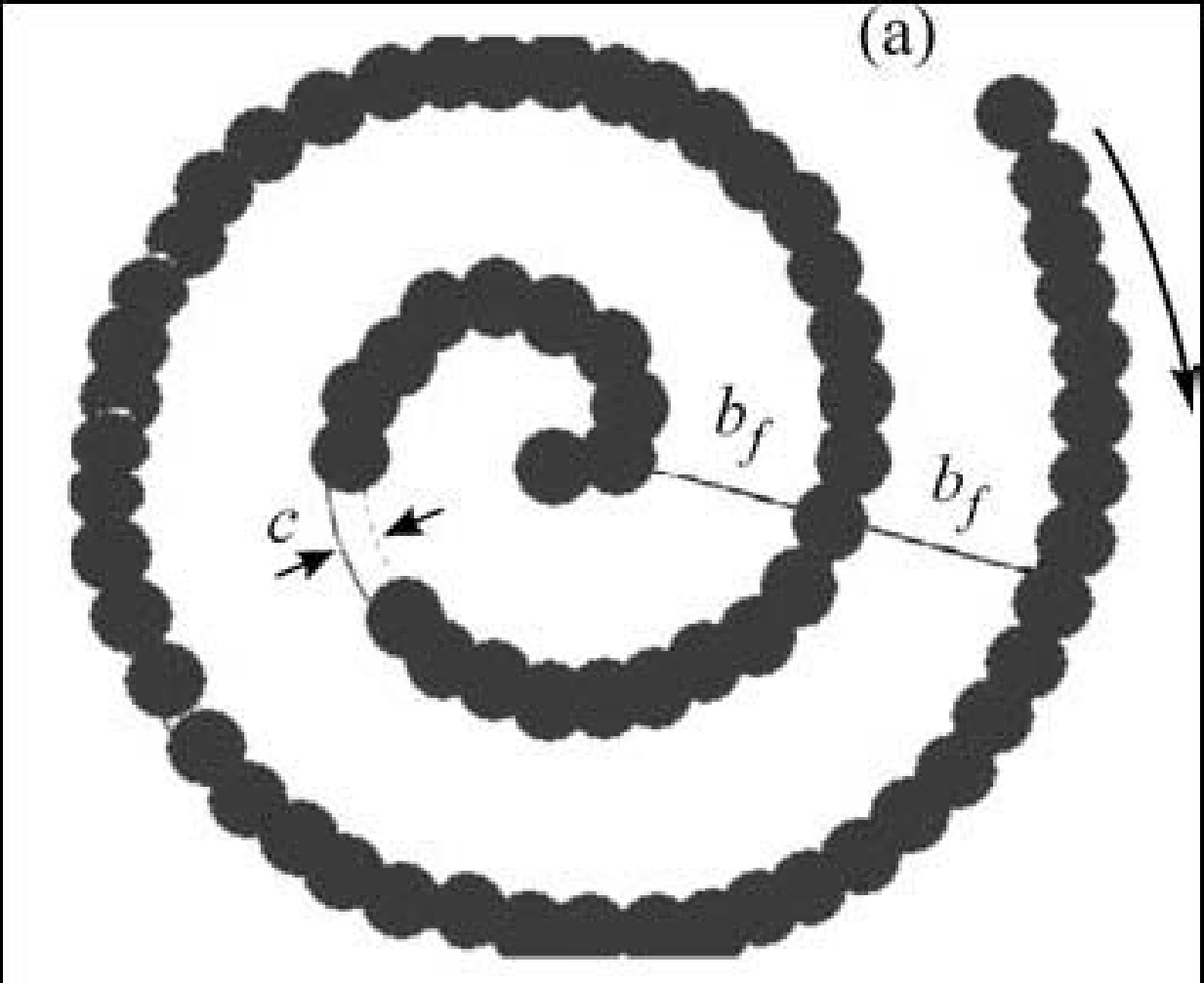}}\vspace{0.0cm}
\centerline{\includegraphics[width=7.5cm,angle=0]{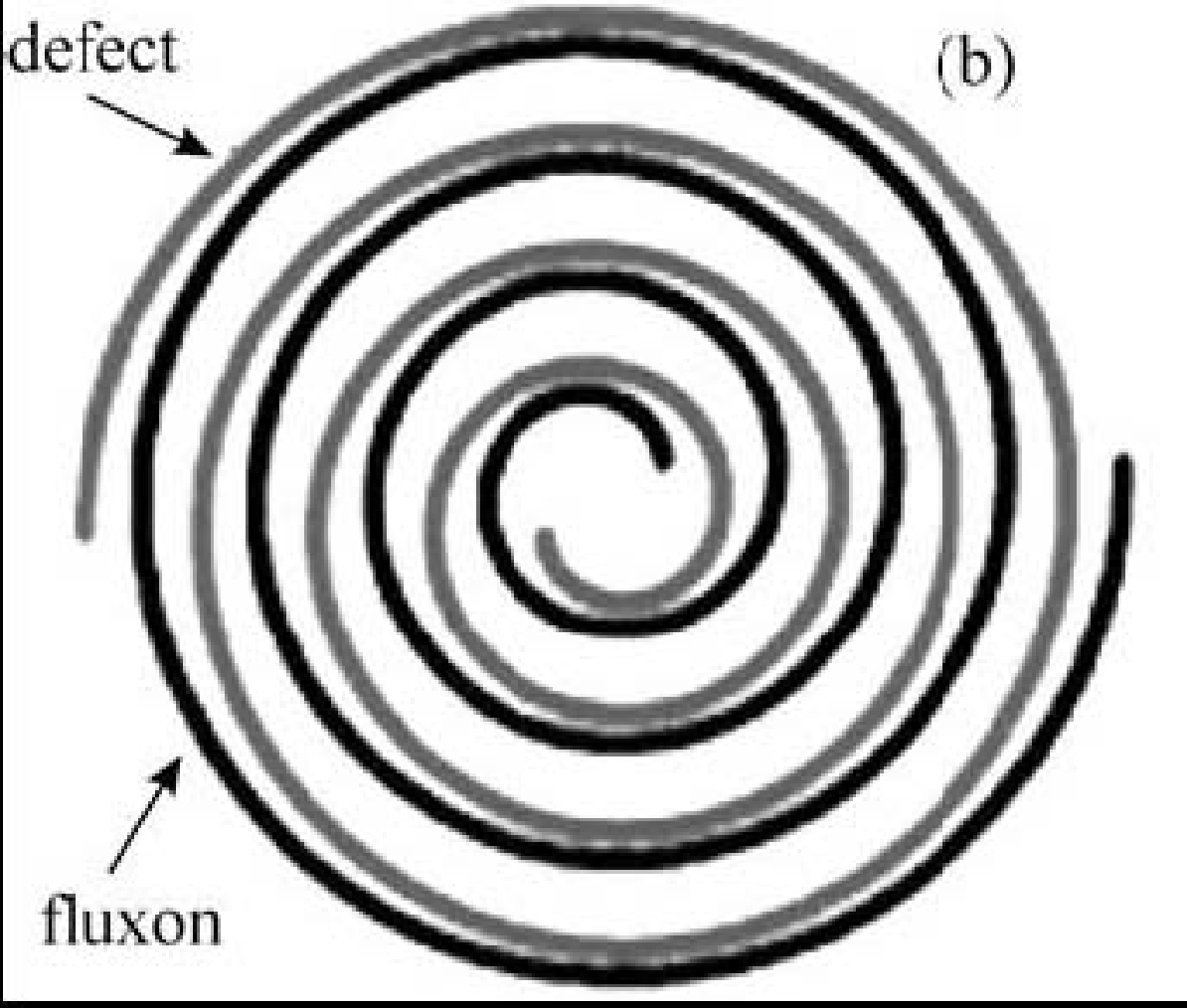}}
\vspace{0.5cm} \caption{Fig.\ref{fig:fig5}. (top) The structure of
the one-armed Josephson spiral fluxon with pitch $b_f$ winding
around the Archimedean spiral of the screw dislocation ($c$ is
distance between terraces along the $c$-axis of $YBCO$ crystal).
(bottom) Trapping (stabilization) of the one-armed Josephson
spiral fluxon by the matching one-armed Archimedean spiral of the
giant screw dislocation.} \label{fig:fig5}
\end{figure}
Above the critical field of the single spiral fluxon $h_f(T)$, the
applied magnetic field first starts to penetrate the weakened
regions (near the surface of the sample) which are situated in the
vicinity of screw dislocations (see Fig.\ref{fig:fig5}). There are
two possibilities depending on the field orientation. Flux can
either enter by climbing up the screw dislocation from bottom to
top (for field applied parallel to the surface) or by sliding down
the screw from top to bottom (for field applied normally to the
surface). In both cases, the freely rotating spiral fluxon gets
trapped (stabilized) inside the Archimedean spiral of the defect
(see Fig.\ref{fig:fig5}).
\begin{figure}
\centerline{\includegraphics[width=8cm]{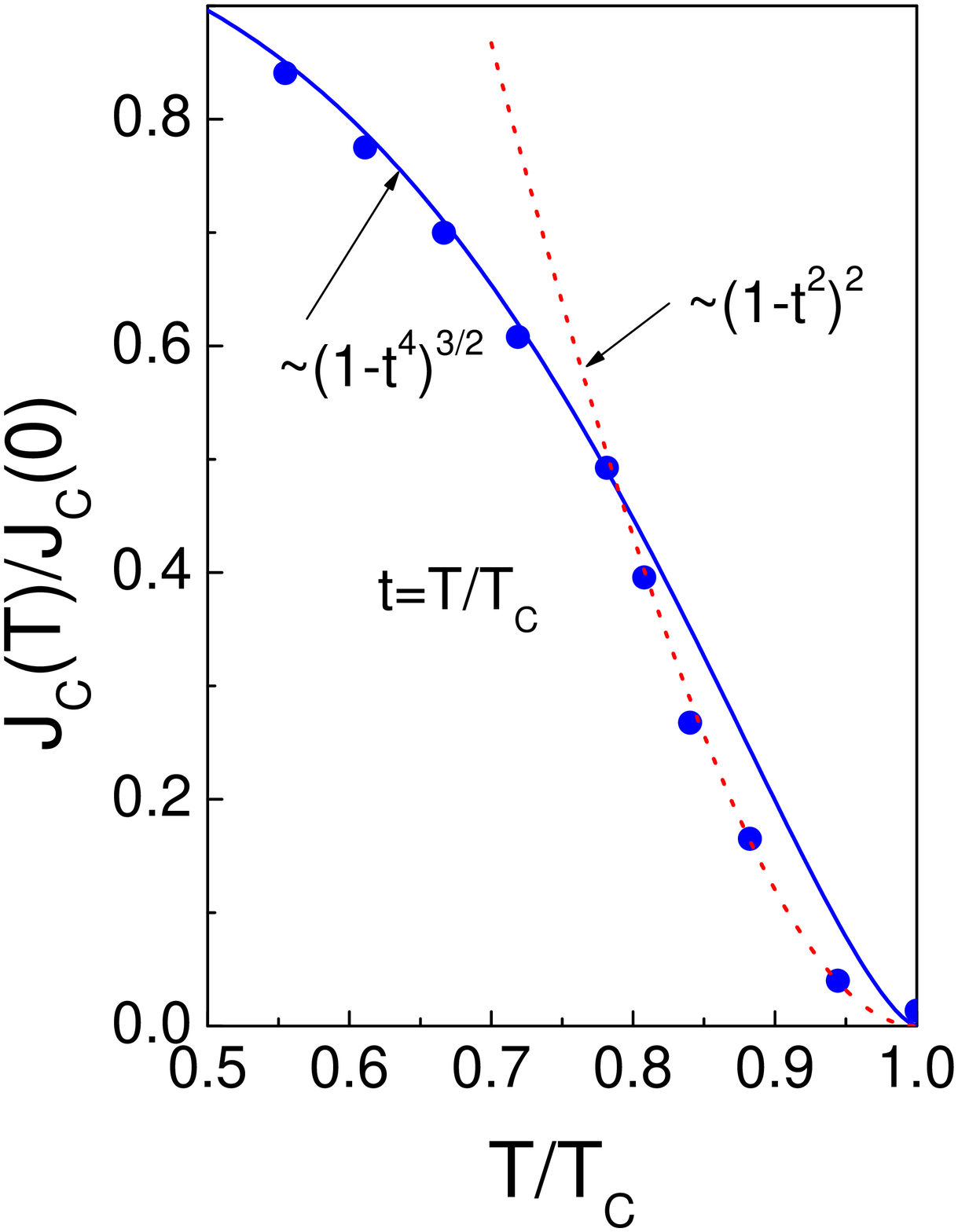}}\vspace{0.5cm}
\caption{Fig.\ref{fig:fig6}. The best fits for the dependence of
the normalized critical current density on reduced temperature
$T/T_C$ deduced from the AC susceptibility $\chi^{'}(T,h_0)$
data.} \label{fig:fig6}
\end{figure}
To discuss the temperature evolution of the introduced parameters,
in Fig.\ref{fig:fig6} we present the fitting results for the
temperature dependence of the normalized critical current density
$J_C(T)$ in our crystals, extracted from the AC susceptibility
data using the usual procedure based on the critical state
model~\cite{19}. A crossover between two different types of
behavior is clearly seen near $T^{*}=0.8T_C$. A low-temperature
region ($T<T^{*}$) was found to follow the well-known~\cite{20}
two-fluid model dependence (solid line in Fig.\ref{fig:fig6})
$J_C(T)=J_C(0)(1-t^4)^{3/2}$ with $t=T/T_C$, while closer to $T_C$
(for $T>T^{*}$) our data are better fitted by a proximity-type
behavior~\cite{20} (dotted line in Fig.\ref{fig:fig6})
$J_C(T)=J_C(0)(1-t^2)^{2}$, most likely related to normal regions
produced by oxygen deficiency and defect related structure. The
extrapolated from the above fits zero-temperature value of the
critical current density in our single crystals is $J_C(0)=5\times
10^8A/m^2$. Assuming~\cite{20} $\lambda_L(0)\simeq 100nm$ for the
London penetration depth and using the above value of $J_C(0)$, we
obtain $\lambda_J(0)\simeq 3.6\mu m$ for an estimate of
zero-temperature Josephson penetration depth which in turn results
in $b_f(0)\simeq 18\mu m/\gamma(0)$ and $h_f(0)\simeq 5.5
\gamma(0)Oe$ for the pitch of the spiral fluxon and the critical
fluxon field, respectively. On the other hand, giant screw
dislocations with a pronounced structure of Archimedean spiral
(with the pitch $b=18.7\mu m$, see Fig.\ref{fig:fig3}),
responsible for creation of Josephson contact areas
$S(T)=db=(2\lambda_L(T)+c)b\simeq 2\lambda_L(T)b$, leads to the
temperature dependence of the peak field $h_p(T)=\Phi_0/S(T)\simeq
\Phi_0/2\lambda_L(T)b$. Using two-fluid model predictions for the
explicit temperature dependence of the London penetration depth
and critical current density, $\lambda_L(T)=
\lambda_L(0)/\sqrt{1-t^4}$ and $J_C(T)=J_C(0)(1-t^4)^{3/2}$ with
$t=T/T_C$, we obtain $\lambda_J(T=50K)\simeq 3.75\mu m$,
$b_f(T=50K)\simeq 18.75 \mu m/\gamma(T=50K)$, $h_f(T=50K)\simeq
5.33\gamma (T=50K) Oe$, and $h_p(T=50K)\simeq 5.34Oe$ for the
Josephson penetration depth, the pitch of the spiral fluxon, the
critical fluxon field, and the peak field, respectively. Observe
that for $\gamma(T=50K)\simeq 1.003$ we have a perfect match
between the screw dislocation spiral pitch $b$ (see
Fig.\ref{fig:fig3}) and fluxon spiral pitch $b_f(T=50K)$ leading
to manifestation of the observed here low-field Josephson FE at
$h_p(T=50K)\simeq h_f(T=50K)$. Likewise, for higher temperatures
the perfect match is achieved using $\lambda_J(T=60K)\simeq
3.95\mu m$ and $\lambda_J(T=70K)\simeq 4.4\mu m$ for the Josephson
penetration depth along with the two fitting parameters
$\gamma(T=60K)\simeq 1.05$ and $\gamma(T=70K)\simeq 1.17$.
Besides, the above-obtained temperature dependence of the peak
field $h_p(T)$ correctly describes the observed slight shifting of
the FE to lower applied fields with increasing the temperature
(namely, $h_p(T=60K)\simeq 5.0Oe$ and $h_p(T=70K)\simeq 4.5Oe$).
Notice also that, for the same temperatures, the lower Josephson
field, $h_J=\Phi_0/\lambda_Jd$, responsible for creation of
conventional $1D$ vortices with the size of $\lambda_J(T)$, is too
high to account for the observed low-field FE. Finally, let us
estimate the pinning force density $f_p$ for a spiral fluxon. By
analogy with Abrikosov vortex core pinning by screw
dislocations~\cite{4,21}, we obtain $f_p\simeq (1/2)\mu_0h_f^2
b_f\simeq 10^{-6}N/m$ which, in the single vortex limit
$f_p=\Phi_0J_C$, corresponds to the maximum value of the critical
current density $J_C\simeq 5\times 10^8A/m^2$ in our crystals.

In summary, employing a highly sensitive homemade AC magnetic
susceptibility technique to single crystals of
$YBa_2Cu_3O_{7-\delta}$ with giant $3D$ screw dislocations (having
the structure of the $2D$ Archimedean spiral), we observed a
manifestation of the low-field analog of the high-field
(Abrikosov) fishtail effect, attributed to highly efficient
pinning (stabilization) of rotating spiral Josephson vortices by
the matching screw defects.

This work has been financially supported by the Brazilian agencies
CNPq, CAPES and FAPESP.

\end{document}